\documentclass[
 reprint,
superscriptaddress,
amsmath,amssymb,
prb,
linenumbers
authblk,
]{revtex4-2}

\usepackage{graphicx} 
\usepackage{dcolumn} 
\usepackage{bm} 
\usepackage{hyperref} 
\usepackage{color}
\usepackage{soul}
\usepackage[dvipsnames]{xcolor}
\usepackage{physics}
\usepackage{ulem} 
\usepackage[utf8]{inputenc}
\usepackage[T1,T2A]{fontenc}
\DeclareUnicodeCharacter{0308}{~}
\graphicspath{ {./} }

\begin{document}

\title{Demonstration of hybrid high-$Q$ hexagonal boron nitride microresonators}

\author{Anustup Das}
\thanks{These authors contributed equally.}
\affiliation{Institute for Quantum Science and Technology, University of Calgary, Calgary, Alberta T2N 1N4, Canada}

\author{Dong Jun Lee}
\thanks{These authors contributed equally.}
\affiliation{Materials Architecturing Research Center, Korea Institute of Science and Technology (KIST), 5 Hwarang-ro 14-gil, Seongbuk-gu, Seoul 02792, Republic of Korea}
\affiliation{Department of Chemical and Biological Engineering, Korea University, 145 Anam-ro, Seongbuk-gu, Seoul, Republic of Korea}

\author{Prasoon K. Shandilya}
\affiliation{Institute for Quantum Science and Technology, University of Calgary, Calgary, Alberta T2N 1N4, Canada}

\author{Sejeong Kim}
\affiliation{School of Mathematical and Physical Sciences, University of Technology Sydney, Ultimo, New South Wales 2007, Australia}
\affiliation{ARC Centre of Excellence for Transformative Meta-Optical Systems (TMOS), University of Technology Sydney, Ultimo, New South Wales 2007, Australia }

\author{Gumin Kang} 
\affiliation{Nanophotonics Research Centre, Korea Institute of Science and Technology (KIST), 5 Hwarang-ro 14-gil, Seongbuk-gu, Seoul 02792, Republic of Korea}

\author{David P. Lake}
\affiliation{Institute for Quantum Science and Technology, University of Calgary, Calgary, Alberta T2N 1N4, Canada}

\author{Bishnupada Behera}
\affiliation{Institute for Quantum Science and Technology, University of Calgary, Calgary, Alberta T2N 1N4, Canada}

\author{Denis Sukachev}
\affiliation{Institute for Quantum Science and Technology, University of Calgary, Calgary, Alberta T2N 1N4, Canada}

\author{Igor Aharonovich}
\affiliation{School of Mathematical and Physical Sciences, University of Technology Sydney, Ultimo, New South Wales 2007, Australia}
\affiliation{ARC Centre of Excellence for Transformative Meta-Optical Systems (TMOS), University of Technology Sydney, Ultimo, New South Wales 2007, Australia }

\author{Jung-Hyun Lee} 
\affiliation{Department of Chemical and Biological Engineering, Korea University, 145 Anam-ro, Seongbuk-gu, Seoul, Republic of Korea}
\email[ASa]{leejhyyy@korea.ac.kr}

\author{Jaehyun Park}
\affiliation{Materials Architecturing Research Center, Korea Institute of Science and Technology (KIST), 5 Hwarang-ro 14-gil, Seongbuk-gu, Seoul 02792, Republic of Korea}
\email{jpark@kist.re.kr}

\author{Paul E. Barclay}
\affiliation{Institute for Quantum Science and Technology, University of Calgary, Calgary, Alberta T2N 1N4, Canada}
\email{pbarclay@ucalgary.ca}

\date{\today}

\begin{abstract}
Hexagonal boron nitride (hBN) is a wide bandgap van der Waals material that is emerging as a powerful platform for quantum optics and nanophotonics. In this work, we demonstrate whispering gallery mode silica microresonators hybridized with thin layers of epitaxially grown hBN that exhibit high optical quality factor $> 7 \times 10^5$. Measurements of the effect of hBN thickness on optical $Q$ and comparison with a theoretical model allows the linear optical absorption coefficient of the hBN films to be estimated. These high-$Q$ devices will be useful for applications in quantum and nonlinear optics, and their hybridized geometry provides a sensitive platform for evaluating losses in hBN and other 2D materials.

\end{abstract}

\maketitle

Two dimensional (2D) materials, in addition to exhibiting many unique physical properties, have potential to be easily integrated with photonic structures due to their natural passivation behaviour stemming from a lack of surface dangling bonds \cite{ref:xia2014two}.
Hexagonal boron nitride (hBN) is a 2D insulating material with an optical bandgap of $\sim 6$ eV, excellent surface morphology, and attractive optical characteristics that are leading to its emergence as a photonics platform for UV to IR wavelengths \cite{caldwell2019photonics}. It can host color center defects that are enabling a wide range of quantum applications \cite{atature2018material}, and it possess enhanced nonlinear \cite{kim2019second, yao2021enhanced} and polaritonic \cite{dai2014tunable, caldwell2014sub} properties.

\begin{figure}[b]
	\includegraphics[width=\linewidth]{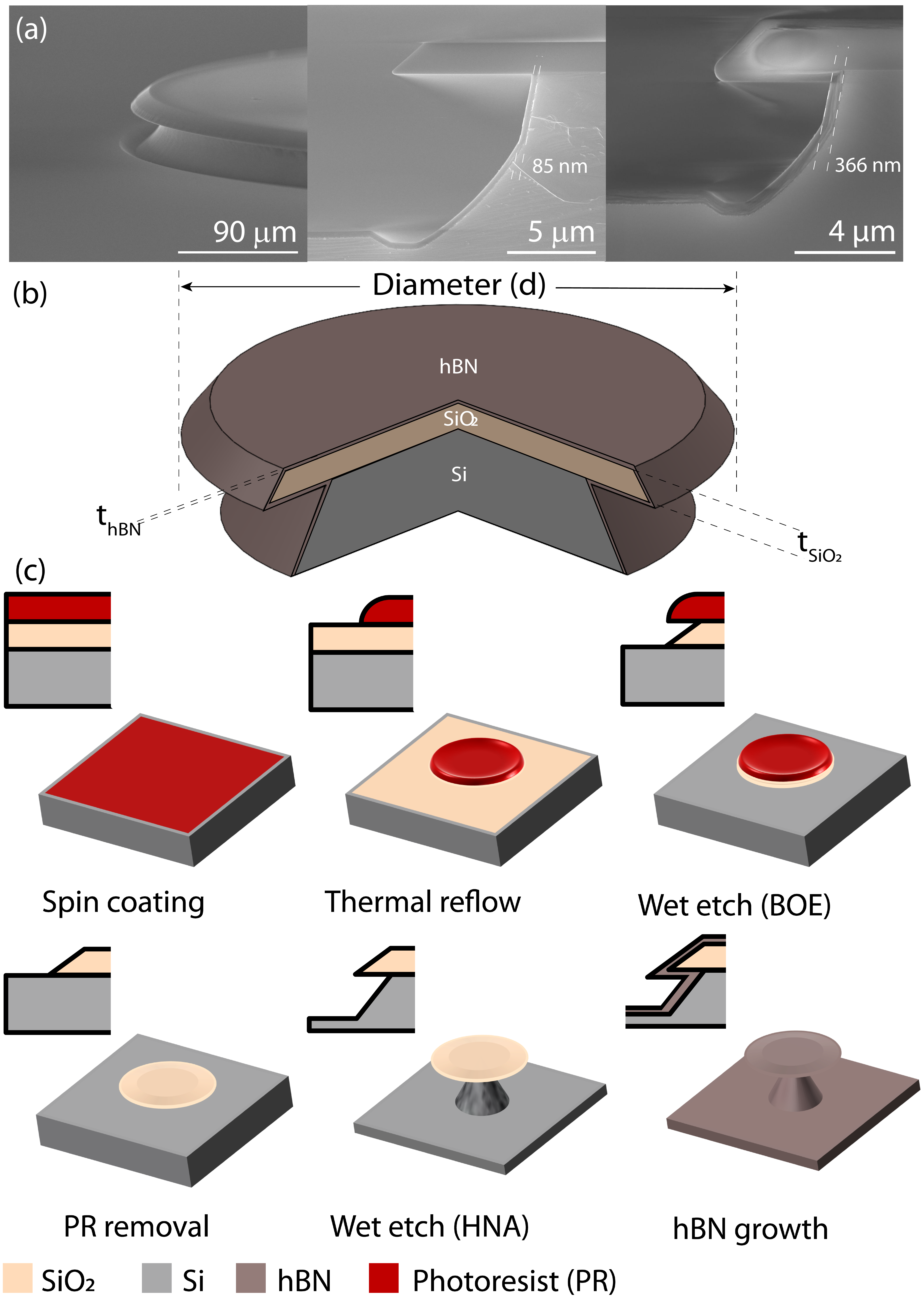}
		\caption{
		\label{fig1} 
		   \textbf{Hybrid hBN-SiO$_2$ device images and fabrication process.}
{(a)} SEM images of the wedged hBN-SiO$_{2}$ microdisk (left) and two line structures, each with different hBN thickness, that were cleaved to reveal the cross-section of the hBN film (center and right). {(b)} Schematic depicting the microdisk structure. {(c)} Illustration of the fabrication process steps.
	    } 
\end{figure}

To fully benefit from hBN's optical and electronic characteristics, it is advantageous to integrate it within on-chip photonic devices such as cavities. Optical microcavities enhance the interaction between light and matter, and have led to fundamental discoveries and applications across many areas of photonics \cite{ref:vahala2003om}.  Microdisk whispering gallery mode microcavities can offer ultrahigh optical quality factor ($Q$) with a relatively low mode volume ($V)$, and have been realized on-chip from a wide range of materials, including wide bandgap materials such as silica \cite{ref:kippenberg2004dem, ref:armani2003uhq}, diamond \cite{Khanaliloo2015, bray2018single}, silicon carbide \cite{Lu2015}, silicon nitride \cite{ref:barclay2006ifc}, aluminum nitride \cite{Sun2019}, lithium niobate \cite{wang2014integrated}, and gallium phosphide \cite{ref:mitchell2014cavity}. 
They have enabled breakthroughs in quantum \cite{ref:peter2005eps, ref:srinivasan2007lno} and nonlinear optics \cite{sandoghdar1996very, ref:spillane2002utr, ilchenko2004nonlinear, ref:kippenberg2004utm, ref:kippenberg2004kno, del2007optical}, optomechanics \cite{ref:kippenberg2005arp, Rokhsari-2005-OpticsExpress-SelfOscillations}, and sensing \cite{ref:vollmer2008wgm}.

Optical microcavities have been integrated with a wide range of 2D materials \cite{Wang:19}. Creating photonic devices from thin layers of hBN has focused primarily on patterning exfoliated material. Photonic crystal cavities \cite{kim2018photonic, froch2019photonic}, microrings \cite{froch2019photonic}, metalenses \cite{liu2018ultrathin}, and bullseye resonators \cite{froch2021coupling} have been patterned from hBN flakes.  Hybrid hBN microcavities have been created by placing hBN flakes over silicon nitride microdisks \cite{ref:Proscia2020hBN} and photonic crystals \cite{froch2020coupling}, and by patterning hBN-tungsten disulfide \cite{ren2018whisper} and other 2D material heterostructures \cite{khelifa2020coupling}.

Fabrication of photonic devices from large areas of hBN grown with controlled thickness and high material quality offers an alternative approach that will enable implementation of scalable 2D material optical technologies. Here, we show that using chemical vapour deposition it is possible to grow uniform thin layers of hBN directly on SiO$_2$ microdisk cavities. The resulting hBN coated devices support whispering gallery modes with intrinsic $Q_i > 7.6 \times 10^5$ (total loaded $Q_t > 5.2 \times 10^5$), nearly three orders of magnitude larger than previous demonstrations \cite{kim2018photonic}. To determine the factor limiting  $Q$ of these devices, the interaction of the microdisk modes with hBN was tuned by varying the hBN thickness between 15 nm and 400 nm. In addition to introducing a platform for creating high-$Q$ hBN photonic devices, these measurements provide insight needed to guide the optimization of future devices and material quality.

Figure \ref{fig1}(a) shows a scanning electron microscope image of a typical hybrid hBN-SiO$_2$ microdisks demonstrated here. As illustrated in Fig.\ \ref{fig1}(b), they consist of a wedged SiO$_2$ microdisk (thickness $t_{\text{SiO}_2} =  1.4\,\mu\text{m})$ supported by a silicon pedestal, and conformally coated with a thin layer of hBN. This hBN layer is visible in the cross-section images in Fig.\ \ref{fig1}(a), which were obtained by cleaving samples patterned with lines in place of microdisks. The fabrication process used to create these devices is illustrated in Fig.\ \ref{fig1}(c). 

Starting with a silica-on-silicon wafer, photolithography (HMDS followed by AZ GXR-601 photoresist) and resist reflow (140 degrees C for 40 minutes) is used to define the SiO$_2$ microdisk pattern. A buffered oxide etch (6:1 BOE, 0.37 $\mu\text{m}$/s etch rate, deionized water rinse) transfers the photoresist pattern into the silica layer, creating a wedged microdisk cross-section \cite{Kang2017}. After removing the photoresist using acetone and isopropyl alcohol, a hydrofluoric, nitric and acetic acid (HNA) wet etch (HF:10 ml, HNO$_{3}$:120 ml, CH$_{3}$COOH:100 ml, 15 min at 25 degrees C) is used to selectively remove the underlying silicon layer, resulting in an undercut SiO$_2$ microdisk supported by a silicon pedestal. 

Chemical vapour deposition is then used to grow a thin layer of hBN on the microdisk structure. This layer is conformal in nature, and its thickness can be varied from between a few nm to several hundred nm by adjusting the deposition time.

The growth process involved bubbling a liquid borazine (B$_{3}$N$_{3}$H$_{6}$) canister maintained at -10 degrees C using N$_{2}$ gas to deliver 20 SCCM of a borazine and N$_{2}$ mixture to a low-pressure CVD system with a quartz tube (outside diameter of 5 cm). An additional 1000 SCCM of H$_{2}$ was supplied as a carrier gas. The thickness of the grown hBN was confirmed by AFM of a uniform large-area pattern formed on an SiO$_2$ wafer. The growth rate was 2.4 nm/min (1 cm × 1cm, thickness uniformity $> 99\%$). The thickness of hBN grown on SiO$_2$ was then color indexed. This color index was used to confirm the thickness of hBN grown on the microdisks by examining the color of the film on the surrounding substrate. Cross-sectional SEM images, such as those shown in Fig.\ \ref{fig1}(a), were used to verify the conformal nature of the hBN growth.

\begin{figure}[t]
	\includegraphics[width=\linewidth]{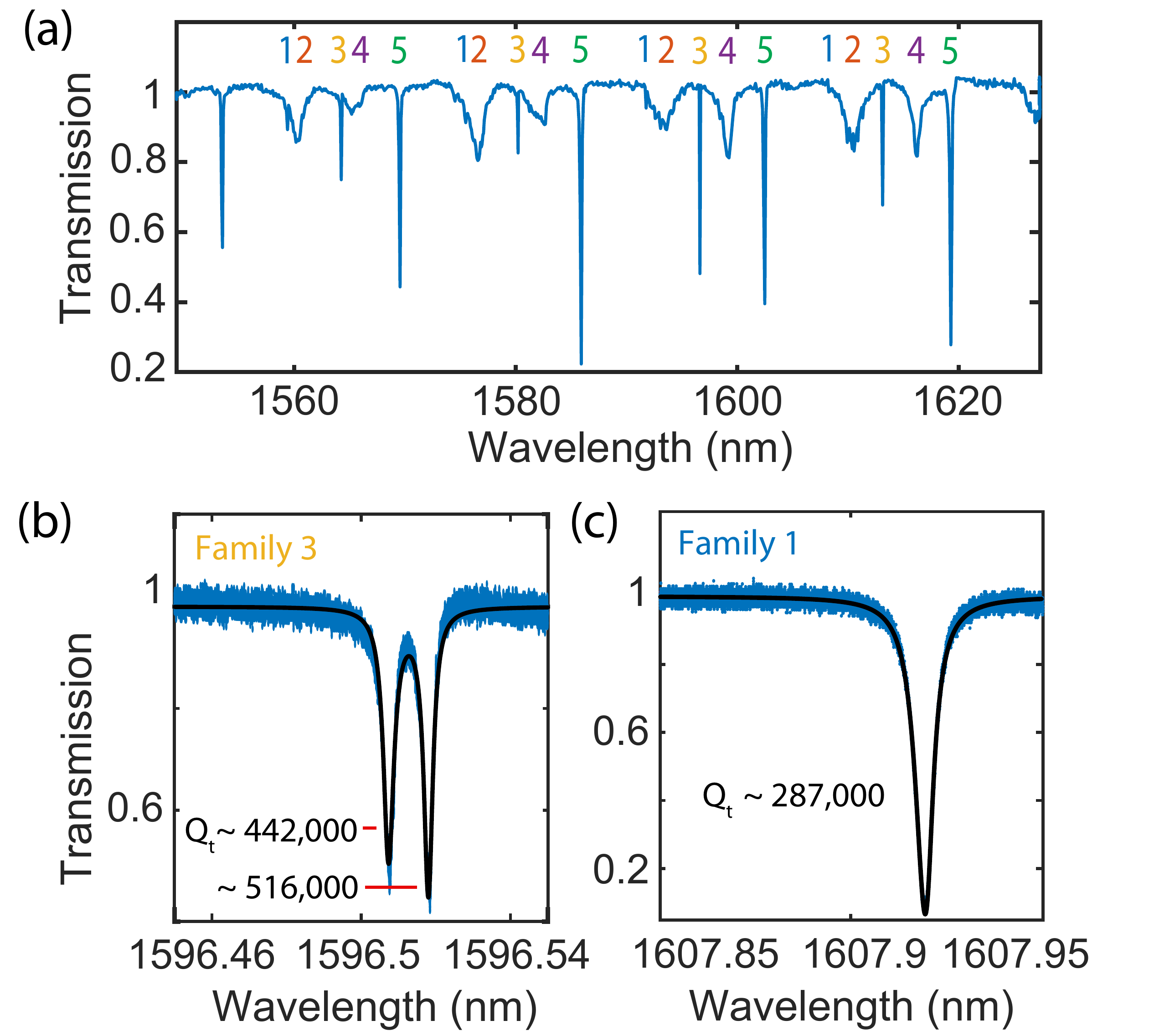}
		\caption{
		    \label{fig2}
		    \textbf{Optical characterization of a microdisk with hBN thickness $t_\text{hBN} = 15\,\text{nm}$ and diameter $d = 30\,\mu\text{m}$.} (a) Broad wavelength scan of the fiber taper transmission in the 1550 nm band when it is coupled to the microdisk. (b,c) Fine wavelength scans of the fiber taper transmission for high-$Q$ modes of the device.
	    }
\end{figure}

To study the optical  properties of the hBN-SiO$_2$ microdisk, an optical fiber taper \cite{ref:birks1992sft} was used to evanescently couple light from tunable diode laser sources into and out of the device. Figure \ref{fig2}(a) shows a typical broad wavelength scan of the fiber transmission in the 1550 nm band (New Focus Velocity laser) when the fiber is coupled to a $d = 30$ $\mu$m diameter device  coated with $t_\text{hBN} = 15$ nm of hBN. The transmission is normalized to the fiber taper transmission when it is far from the microdisk, and the measurement was made with the taper hovering in the near-field of the microdisk. Regularly spaced sharp dips in the transmission correspond to coupling to the device's whispering gallery modes. Families of modes with similar spacing, linewidth, and resonance contrast can be identified. Within each family of modes the azimuthal mode number $m$ varies, while the radial and vertical mode quantum numbers are constant. For the measurements in Fig.\ \ref{fig2}(a), the polarization of the input field was optimized to maximize contrast of mode family 2.

A high resolution scan of a resonance from the highest $Q$  mode family 3 is shown in Fig.\ \ref{fig2}(b). This measurement was taken using the piezo scanning capabilities of the laser source, and with the input field polarization optimized to maximize the resonance contrast.  It reveals that the resonance is a doublet, resulting from backscattering from roughness or other defects within the device that couple degenerate clockwise and counterclockwise propagating modes to create non-degenerate symmetric and anti-symmetric standing wave modes \cite{ref:borselli2005brs}. The higher-$Q$ standing wave mode of the doublet possesses a total loaded $Q_t = 5.2 \times 10^5$. From the resonance contrast we find the mode's intrinsic $Q_i = 7.8 \times 10^5$, where we have taken into account the standing wave mode's equal coupling to both the forward and backward propagating fiber taper modes \cite{SpillaneSMPainterOJandVahalaKJ2003,ref:borselli2005brs}, resulting in the relationship $Q_i = Q_t/\sqrt{T_o}$.  Figure \ref{fig2}(c) shows a fine scan of a mode from the next highest-$Q$ mode family 1, which has $Q_t = 2.9\times 10^5$. This mode is a singlet, as are other lower-$Q$ mode families for this device whose higher internal loss rate exceeds the mode coupling rate from back-scattering.

To identify each family of resonances, the modes of the microdisk were calculated numerically using finite difference time domain (FDTD) simulations. Based on each family's measured free spectral range and its dependence on wavelength, we identify mode families 1 and 3 as the fundamental TE-like and  TM-like modes, respectively.  From FDTD we predict that the measured $Q$ for both of these mode families is not limited by radiation loss. As discussed in more detail below, the TE-like mode is predicted to more strongly overlap with the hBN layer. This, combined with the fact that its $Q_t$ is smaller than that of the TM-like mode suggests that the hBN is contributing to optical absorption in these devices.

To study the influence of the hBN on the microdisk $Q$, we characterized devices with varying hBN thickness. For these measurements we used large diameter $d = 180\,\mu\text{m}$ microdisks to ensure that radiation loss was playing a minimal role. 

\begin{figure}[tb]
	\includegraphics[width=\linewidth]{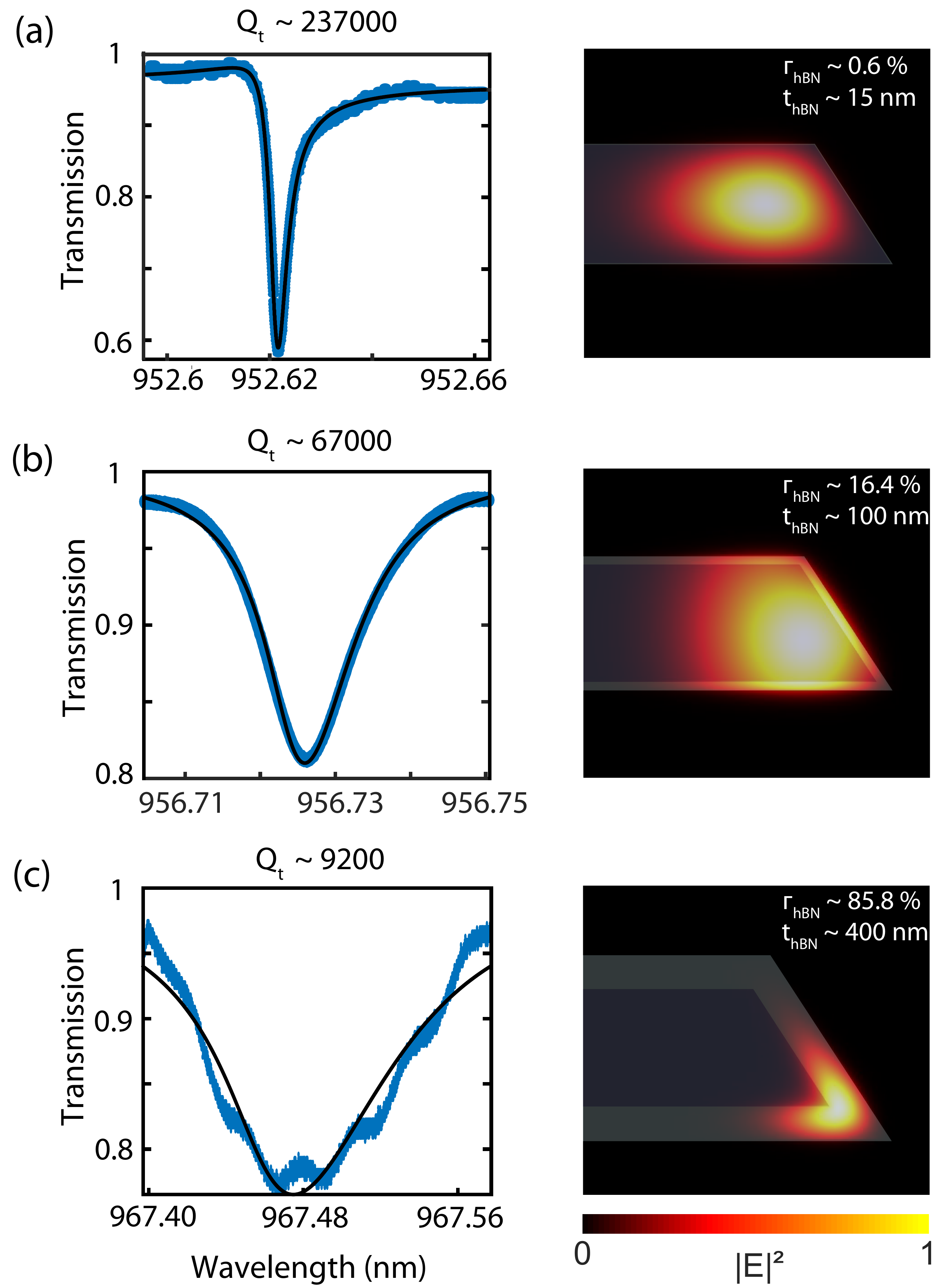}
		\caption{
		    \label{fig3}
		   \textbf{Characterization of optical modes for varying hBN thickness.} Narrow wavelength scans (left) and FDTD simulations (right) of the highest $Q$ modes of $d = 180\,\mu\text{m}$ devices with $t_\text{hBN} = [15, 100, 400]\,\text{nm}$ are shown in (a) -- (c), respectively. 
		   }
\end{figure}

\begin{figure}[t!]
	\includegraphics[width=\linewidth]{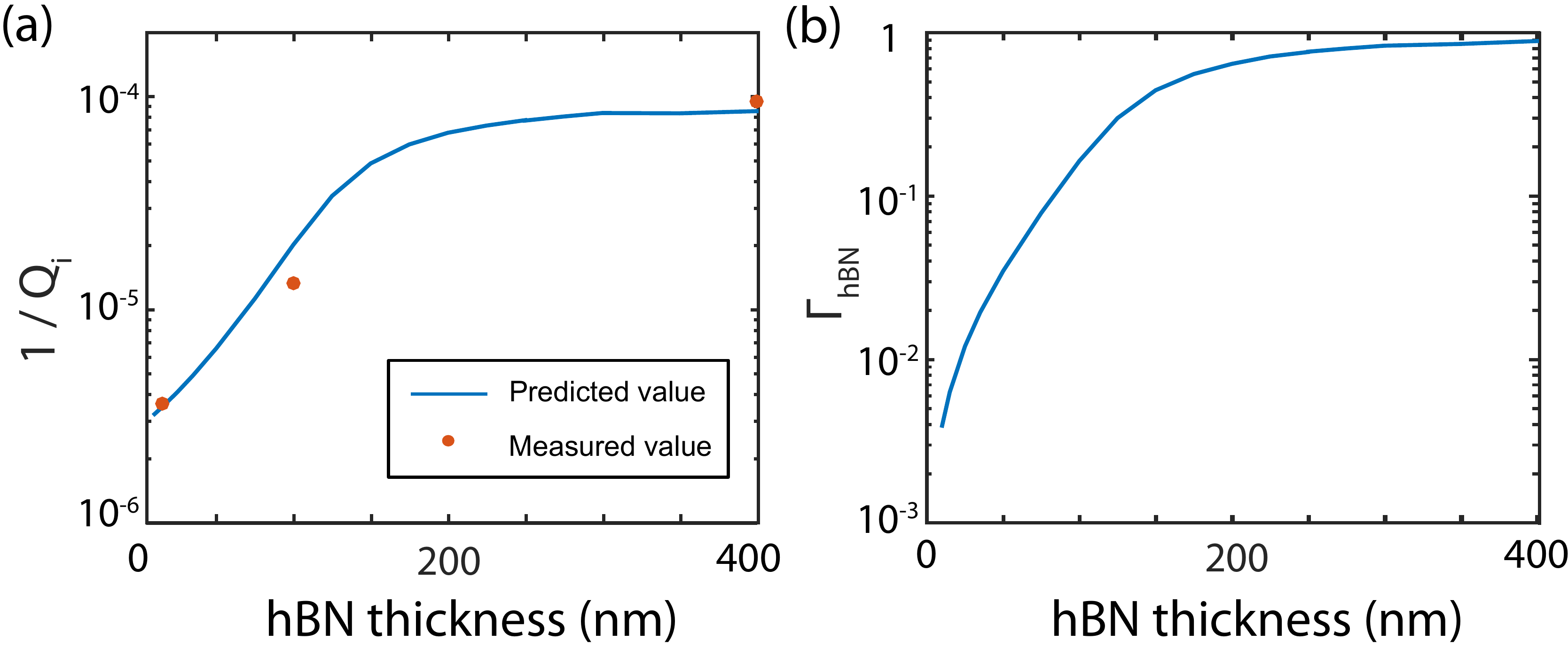}
		\caption{
		    \label{fig4}
{\textbf{Dependence of microdisk loss on hBN layer thickness.}} (a) Measured (blue points) and simulated  (red points) intrinsic decay rate {and (b) confinement factor ($\Gamma_\text{hBN}$)}, for varying hBN thickness. Measurements were made in the 950 nm wavelength band. Simulations of $\Gamma_\text{hBN}$ considered the fundamental TE-like mode. {The model of $1/Q_i$ assumes that the hBN has linear absorption coefficient $\alpha_\text{hBN} = 9.5~\text{cm}^{-1}$}. 
}
\end{figure}

This is investigated in more detail in Fig.\ \ref{fig3}, which shows the highest measured $Q$ mode of three devices that have different hBN thicknesses $t_\text{hBN} = \left[15, 100, 400\right]\,\text{nm}$ but are otherwise nominally identical.  From Fig.\ \ref{fig3}, it is immediately apparent that $Q$ is strongly reduced by increasing $t_\text{hBN}$, falling from $Q_t = 2.4\times10^5$ for $t = 15\,\text{nm}$ (Fig.\ \ref{fig3}(a)) to $Q_t = 6.7\times10^4$ and $9.2\times10^3$ for $t_\text{hBN} = 100$ and $400$ nm, respectively (Figs.\ \ref{fig3}b and \ref{fig3}c). Figure \ref{fig4}(a) plots the corresponding measured $1/Q_i$, which is proportional to the intrinsic photon decay rate, as a function on $t_\text{hBN}$. 

These measurements were made using a tunable laser in the 950 nm band,  which was found to couple more efficiently to the microdisks with thicker hBN owing to the higher effective index of the fiber taper waveguide mode at shorter wavelengths. The larger effective index provides better phase matching to microdisk modes significantly confined in the hBN, whose index of refractive exceeds that of silica. Phase matching is particularly important when coupling to large diameter devices such as these. In future it will be possible to optimize the fiber diameter for each hBN thickness to ensure optimal phase matching at 1550 nm. 
The 950 nm wavelength range is also of interest given its proximity to the emission wavelengths of typical hBN colour centers (570 - 820 nm). 
{Finally, at shorter wavelengths it is possible to more strongly confine the optical field in the hBN, providing a more sensitive measure of its optical properties. }

The strong dependence of $Q_t$ on $t_\text{hBN}$ can be explained by the increase in overlap of the microdisk mode with the hBN layer as the film thickness increases. This overlap is quantified by the confinement factor,
\begin{equation}
\Gamma_\text{hBN} = \frac{\int_\text{hBN}d\textbf{r}\, n_\text{hBN}^2|E(\textbf{r})|^2}{\int d\textbf{r}\, n^2(\textbf{r})|E(\textbf{r})|^2}, \label{eq:Gamma}
\end{equation}
which describes the fraction of a microdisk mode's energy stored within the hBN. Here $|E(\textbf{r})|$ is the optical mode field amplitude as a function of position $\textbf{r}$, $n(\textbf{r})$ describes the refractive index profile of the microdisk and its surrounding volume, and $n_\text{hBN} \sim 1.84$ is the refractive index of hBN in the near-infrared spectrum. The integral in the numerator of Eq.\ \eqref{eq:Gamma} is evaluated over the hBN region. The confinement factor can be related to the loss rate due to absorption from the hBN by:
\begin{equation}
\gamma_\text{hBN} = \Gamma_\text{hBN} \alpha_\text{hBN} {v_g}, \label{eq:alpha}
\end{equation}
where $\alpha_\text{hBN}$ is the linear absorption coefficient of the hBN material, in units of inverse meters, and $v_g$ is the group velocity of the optical mode of interest \cite{ref:borselli2005brs}. This shows that as confinement increases, we expect an increase in optical absorption rate. 

In Fig.\ \ref{fig3} we show the field profiles and corresponding $\Gamma_\text{hBN}$ of the fundamental TE-like mode of the microdisk for varying $t_\text{hBN}$, as calculated using finite element  simulations (COMSOL). These show that as $t_\text{hBN}$ increases, the mode becomes  more confined within the hBN layer, transforming from a mode that weakly overlaps with the hBN layer for 15 nm thickness, to a mode that is dominantly confined within the hBN for 400 nm thickness. This effect is in part due to the higher refractive index of hBN compared to SiO$_2$, which allows the thicker hBN to support highly confined index-guided modes.

To evaluate the whether this dependence of $\Gamma_\text{hBN}$ on hBN thickness can explain the observed behaviour of $Q_t$, we model the total intrinsic loss of the device as $\gamma_{i} = \gamma_\text{hBN} + \gamma_{i}^o$, where $\gamma_{i}^o$ includes other loss mechanisms such as surface roughness and absorption, loss within the SiO$_2$ microdisk, and radiation loss. Here we take it to be  independent of hBN thickness, though in general it can vary due to both the microdisk field profile's dependence on $t_\text{hBN}$ and changes in the physical properties of the hBN film with varying thickness. A comparison between the measurements and our model is shown in Fig.\ \ref{fig4}(a), exhibiting good agreement between experiment and theory.

Here {$\alpha_\text{hBN} =  9.5\,\text{cm}^{-1}$} and $Q_{i}^o = 6.0\times10^5$ ($\gamma_{i}^o = \omega/Q_{i}^o$) are the sole fitting parameters. The simulated  $\Gamma_\text{hBN}(t_\text{hBN})$ used to generate this fit was obtained for the fundamental TE-like mode, and is plotted in Fig.\ \ref{fig4}(b).

Deviations between the predicted and observed $\gamma_i(t_\text{hBN})$, as well as errors in our estimates of $\alpha_\text{hBN}$, can arise from a number of sources. When measuring modes in large diameter microdisks, it can be challenging to identify measured resonances with specific modes of the device, owing to the large number of high-$Q$ mode families supported by the microdisk. However, other modes are predicted to have similar or smaller confinement than the fundamental TE-like mode, and their qualitative dependence of $\Gamma_\text{hBN}$ on thickness is similar. For example, the fundamental TM-like mode has $\Gamma_\text{hBN} = \left[0.0046, 0.11, 0.89\right]$ for $t_\text{hBN} = \left[50,100,400\right]~\text{nm}$, i.e.\ of a similar order of magnitude and with similar $t_\text{hBN}$ dependence as for the fundamental TE-1 mode shown in Fig.\ \ref{fig3}.

Additional sources of error can arise from other aforementioned contributions to microdisk optical loss, such as surface scattering, which may depend on hBN thickness. Variations in processing and material growth conditions between devices could also result in deviations in $Q_{i}$ not accounted for in our model. 
In future, measurements of the dependence of $Q_{i}$ on wavelength or mode order for a given device will provide further insight into the loss mechanisms without uncertainty related to variations between devices. Power dependent measurements may also allow separation of scattering and absorption related contributions to $Q_{i}$ \cite{wang2021using}.

In summary, we have demonstrated a hybrid hBN-SiO$_2$ microdisk, and have characterized the effect of hBN on the device performance. The devices demonstrated here will be of immediate interest to experiments involving hBN optomechanics \cite{shandilya2019hexagonal} and nonlinear optics \cite{kim2018photonic, yao2021enhanced}.

Compared to previously studied hBN cavities \cite{kim2018photonic, ref:Proscia2020hBN, ren2018whisper},

these devices have orders of magnitude higher-$Q$.

Future work will aim to decrease their mode volume while maintaining or increasing $Q$, in order to further enhance light-matter interactions central to these applications. Simulations indicate that by decreasing $d$ to $25\,\mu\text{m}$, we can reduce the mode volume of our high-$Q$ device to $V = 110 (\lambda/n)^3$, as defined by the peak electric field intensity, while maintaining a radiation loss limited $Q > 10^6$ at $1550\,\text{nm}$. Further decreasing $V$ through use of a thicker hBN layer, as in the device in Fig.\ \ref{fig3}(c) is also possible thanks to hBN's higher refractive index compared to SiO$_2$, with $V \sim 80 (\lambda/n)^3$ achievable in a $t_\text{hBN} = 400\,\text{nm}, d = 20\,\mu\text{m}$ microdisk. However, to maintain high-$Q$ in these devices it will be necessary to decrease $\alpha_\text{hBN}$  by {ensuring continuous and defect free hBN growth over the device profile}. The microdisk platform demonstrated here is ideally suited for sensitively probing material quality, and will aid efforts to understand and optimize hBN growth conditions.

%

\end{document}